\documentclass[12pt,preprint]{aastex}

\begin{document}

\title{Rayleigh Scattering and Microwave Background Fluctuations}

\author{Qingjuan Yu, David N. Spergel\footnote{W.M. Keck Distinguished
Visiting Professor of Astrophysics, Institute for Advanced Study,
Princeton, NJ 08540} \& Jeremiah P. Ostriker \\
Department of
Astrophysical Sciences, Princeton University, Princeton, NJ 08544}

\begin{abstract}
During the recombination epoch, cosmic background photons couple not only to
free electrons through Thompson scattering, but also to the neutral hydrogen
through Rayleigh scattering. This latter is $\sim 2$\% effect for
photons near the peak of the photon energy distribution at $z=800$
and a $\sim 0.2\%$ effect at $z=1100$. Including Rayleigh scattering
in the calculation reduces Silk damping at fixed redshift, alters
the position of the surface of last scattering and alters
the propagation of acoustic waves.  We estimate the amplitude
of these effects. For the {\it Microwave Anisotropy Probe} ({\it MAP}),
Rayleigh scattering increases the anisotropy
spectrum by 0.1\% at the most.
For the highest frequencies of the {\it Planck Surveyor},
the effects of Rayleigh scattering are much more dramatic
(decreasing the anisotropy spectrum by 3\% at $\nu\sim$ 550GHz and $l\sim
1000$). The relative difference between the spectra of 
low and high frequencies is imposed by an oscillation with a function of
multipole $l$ and the oscillation amplitude is up to 0.5\% between 100 and
550~GHz.
Rayleigh scattering also slows the decoupling between radiation and matter,
but the effect is undetectably small.

\end{abstract}

\section{Introduction}

With the upcoming launch of the {\it Microwave Anisotropy Probe}
({\it MAP})\footnote{%
For details, see http://map.gsfc.nasa.gov.} in 2001 and the {\it Planck Surveyor}\footnote{For details, see http://astro.estec.esa.nl/Planck/.} in 2007, the
study of the cosmic microwave background (CMB) fluctuations is about to enter
a new epoch. These
satellites\ will characterize the statistical properties of the microwave
background at better than the 1\% level. This dramatic improvement in
experimental capabilities demands that theorists attempt to achieve this
level of accuracy in their calculations. Otherwise, systematic errors in the
theoretical calculations will lead to systematic errors in our
interpretation.  Over the past several years, several 
corrections to the CMB spectrum have been
identified that systematically
alter the predictions of a given model at the 1\% level
\citep{HSSW,GG,SSS99}.  Without these corrections, theoretical
systematic biases could be as important as experimental error.
This paper explores another possible source of systematic error:
neglecting Rayleigh scattering in estimates of the CMB fluctuation
spectrum.

In their seminal paper, \citet{PY} note that Rayleigh scattering
accounts for a few percent of the opacity near decoupling. To simplify
their analysis, they ignore it and include only Thompson scattering in their
calculation. Most subsequent analytical and numerical work has also included
only Thompson scattering. This approximation makes the amplitude of cosmic
microwave background fluctuations a simple function of frequency and greatly
simplifies the equations. \citet{H01} considers the effect of Rayleigh
scattering; however, since he does not do a frequency-dependent calculation,
he significantly underestimates its importance.

In this paper, we estimate the effects of Rayleigh scattering on microwave
background fluctuations. Our calculation begins with the analytical approach of
\citet{P} and \citet{HS95}: we treat the photon-baryon fluid
as tightly coupled through the surface of last scattering. While this approach
is not exact, it enables us to estimate and identify the effects of Rayleigh
scattering. In section 2, we discuss the propagation of acoustic waves in
the baryon-photon fluid. The additional opacity source reduces photon
diffusion and also leads to a frequency dependence in the phase of the
acoustic wave. In section 3, we identify four different observational
effects. While some of these effects are important only at high frequencies,
the suppression of Silk damping leads to observationally significant changes
in the Rayleigh-Jeans regime. In section 4, we check our analytical
estimates with a full numerical  integration of
the full Boltzmann equations using a modified version of CMBFAST.
Section 5 concludes. In this paper, if not specified, the speed of light is
set to be 1.

\section{Acoustic Wave Propagation}

Rayleigh scattering adds an additional source
of opacity to photon propagation:
\begin{equation}
\lambda^{-1}(\nu )=n_e\sigma _T+n_H\sigma _T\left({\nu \over \nu_{eff}}\right)^4,
\label{eq:lambda}
\end{equation}
where $n_e$ is the electron density, $n_H$ is the neutral hydrogen density, $%
\sigma _T$ is the Thompson cross-section, and
\begin{equation}
\nu_{eff}^{-2}\equiv\sum_{j=2}^{\infty}f_{1j}/(\nu_{1j}^2-\nu^2)\simeq\sum_{j=2}^{\infty}f_{1j} \nu_{1j}^{-2}\simeq(0.95 c R_A)^{-2}, \qquad {\rm for}\ \  \nu\ll\nu_{12},
\end{equation}
where $f_{1j}$ is the Lyman series oscillator strength, $\nu_{1j}$
is the Lyman series frequency, $c$ is the speed of light and $R_A$ is the
Rydberg constant of hydrogen atoms
(see Lang 1999, equation [1.306], and the oscillator strength in
equation [2.118] and references therein).
Looking at equation (\ref{eq:lambda}) and ignoring the frequency dependence,
one sees that, roughly speaking, addition of Rayleigh scattering corresponds
to an increasing $n_e$ or $\Omega_b h^2$ (where $\Omega_b$ is the baryon density
in units of critical density and the Hubble constant
$H_0=100h~{\rm km~s^{-1}~Mpc^{-1}}$), which will be seen to be true at low
frequencies (i.e., the increasing of anisotropy spectra at $\nu\la 150$GHz)
in \S 4.
Other atomic and molecular processes are unimportant: the abundances of H%
$_2$, HD and other molecules are low in the early universe \citep{SLD}
and neither He or H$^{-}$ are significant sources of
opacity at the relevant frequencies.  The fractional abundance of hydrogen
in the $n=2$ state peaks at 10$^{-13}$ just prior to decoupling \citep{SLD},
so Balmer lines are not a significant source of
opacity.

In this section, we explore the effects of a {\it frequency dependent}
cross-section on acoustic wave propagation.
Assuming that the radiation is unpolarized,
we begin by expanding the microwave background fluctuations into a smooth
unperturbed piece $f^{(0)}(\nu )=[\exp (h_{\rm P}\nu/k_{\rm B}T)-1]^{-1}$
(where $h_{\rm P}$ is the Planck constant, $k_{\rm B}$ is the Boltzmann constant,
$\nu$ is the
comoving frequency and $T$ is the present temperature), and, in the
conformal Newtonian gauge, a perturbation
piece that obeys the linear collisional Boltzmann equation 
\citep{WS,BE,DJ,HS95}:

\begin{eqnarray}
{\partial f^{(1)}(\nu,k,\mu)\over\partial \eta} 
+ik\mu \bigg[f^{(1)}(\nu,k,\mu)-\frac{\partial f^{(0)}(\nu)}{\partial\ln\nu}\Psi(k,\eta)\bigg]=
\frac{\partial f^{(0)}(\nu)}{\partial\ln\nu} {\partial \Phi(k,\eta) \over \partial \eta}
+\frac{a}{a_0}\frac{1}{\lambda[(1+z(\eta))\nu]}\nonumber \\
\times\bigg[f^{(1)}_0(\nu,k)+\frac{1}{2}f^{(1)}_2(\nu,k)P_2(\mu)-f^{(1)}(\nu,k,\mu)
-\frac{\partial f^{(0)}(\nu)}{\partial\ln\nu} \mu u_b(k)\bigg],
\label{eq:Boltzmann}
\end{eqnarray}
where $\mu={\bf\hat{k}\cdot\hat{n}}$, ${\bf\hat{k}}$ is the comoving
wavevector of the perturbation, ${\bf\hat{n}}$ is the unit vector in the
direction of propagation, $k=|\bf\hat{k}|$ is the comoving wavenumber,
$u_b$ is the baryon velocity, $\eta$
is the conformal time and $z(\eta$) is the redshift at $\eta$. 
The $a$ is the scale factor and its value at the present time is $a_0$.
$\Psi$ and $\Phi$ are the Newtonian potential and the perturbation to the 
intrinsic spatial curvature.
The photon distribution function is expanded in Legendre polynomials:
\begin{equation}
f^{(1)}(\nu ,k,\mu )=\sum_{l=0}f_l^{(1)}(\nu ,k)(2l+1)P_l\left( \mu \right).
\end{equation}
Ignoring changes in $R$(=$3\rho_\gamma/4\rho_b$, where $\rho_\gamma$ and
$\rho_b$ are the radiation density and baryon density respectively) and terms
involving $\Phi$ and $\Psi$, a moment expansion
of equation (\ref{eq:Boltzmann}) gives (e.g., Hu \& Sugiyama 1996): 
\begin{eqnarray}
\dot f_0^{(1)}(\nu ,k) &=&-ikf_1^{(1)}(\nu ,k),
\nonumber \\
\dot f_1^{(1)}(\nu ,k) &=&-\frac{ik}3\left[
f_0^{(1)}(\nu ,k)+2f_2^{(1)}(\nu ,k)\right] -\frac{a}{a_0}\frac 1{\lambda[(1+z(\eta))\nu]}\left[
f_1^{(1)}(\nu ,k)+\frac 13\frac{\partial f^{(0)}(\nu )}{\partial \ln \nu }%
u_b(k)\right],  \nonumber \\
\dot f_2^{(1)}(\nu ,k) &=&-\frac
 25ikf_1^{(1)}(\nu ,k)-\frac{a}{a_0}
\frac 1{\lambda[(1+z(\eta))\nu]}\left[ f_2^{(1)}(\nu ,k
)\right],  \label{LBexpansion}
\end{eqnarray}
where we have truncated the expansion beyond the second moment and ignored 
the angular dependence of Compton and Rayleigh scattering.
Overdots represent derivatives
to the conformal time $\eta$.
The baryon velocity obeys the familiar mass and momentum conservation
equations:
\begin{eqnarray}
\dot\delta _b(k) &=&iku_b(k),  \label{velocity} \\
\rho _b\dot u_b(k) &=&\frac{a}{a_0}\int\frac{8\pi h_{\rm P}\nu ^3d\nu }{%
\lambda[(1+z(\eta))\nu]}\left[ f_1^{(1)}(\nu ,k)+\frac{\partial f^{(0)}(\nu )}{%
\partial \ln \nu }\frac{u_b(k)}3\right] -i\rho _bc^2_bk\delta _b(k),  \nonumber
\end{eqnarray}
where $\delta_b$ is the baryon over-density and $c_b$ is the baryon sound speed.
This equation explicitly allows for the extra photon drag due to Compton and
Rayleigh scattering, and, consequently, for the slightly later momentum
decoupling and lower amplitude of the baryon fluctuations. But the heating on
the baryon due to the extra photon drag of Rayleigh scattering can be neglected
because of the small relevant ratio for comparing Rayleigh and Compton heating 
$\sim(m_e/m_p)[\langle\lambda^{T}/\lambda(\nu)\rangle-1]\sim 10^{-5}$,
where $m_e/m_p$ is the electron mass to proton mass ratio,
$\langle\lambda^{T}/\lambda(\nu)\rangle$(=$\lambda^T/\tilde\lambda(\eta)$, see
equation 7) is the mean free path ratio of photons
without Rayleigh scattering to photons with Rayleigh scattering,
which can be estimated from Figure~\ref{fig:visi} later.
We assume that the baryon sound velocity is small and ignore its pressure.

We define a momentum weighted mean free path, which is equivalent to the
Rosseland mean opacity in stellar atmospheres:
\begin{eqnarray}
\frac 1{\tilde \lambda(\eta)} &=&\int \frac{\nu ^3d\nu }{\lambda[(1+z(\eta))\nu]}\frac{%
\partial f^{(0)}(\nu )}{\partial \ln \nu }\left/ \int \nu ^3d\nu \frac{%
\partial f^{(0)}(\nu )}{\partial \ln \nu }\right.   \nonumber \\
&=&-\frac 1{4\rho _\gamma }\int \frac{8\pi h_{\rm P}\nu^3d\nu }{\lambda[(1+z(\eta))\nu]}\frac{%
\partial f^{(0)}(\nu )}{\partial \ln \nu }.
\end{eqnarray}
Following \citet{P}, we look for solutions proportional to exp$\left(
i\int \omega (k,\eta )\right) .$ We expand the solutions in $\delta
\lambda =(\lambda[(1+z(\eta))\nu]-\tilde \lambda(\eta))${
\begin{eqnarray}
f_1^{(1)}(\nu ,k)=&&A(k)\frac{\partial f^{(0)}(\nu )}{\partial \ln \nu }[
1+b(k)\delta \lambda +o(\delta\lambda)]
\exp\left[ -\int i\omega(k,\eta) d\eta\right].
\end{eqnarray}

Substituting this back into equation (\ref{eq:Boltzmann}), we find
\begin{eqnarray}
f_0^{(1)}(\nu ,k) &=&\frac{-k}{\omega (k)}f_1^{(1)}(\nu ,k);  \nonumber \\
f_2^{(1)}(\nu ,k) &=&-\frac 25ik\lambda[(1+z(\eta))\nu]f_1^{(1)}(\nu ,k);
\label{expand}
\end{eqnarray}
and (\ref{velocity}):
\begin{equation}
u_b(k)=-3A(k)\frac{1}{1-i\omega (k)R\tilde\lambda}\exp\left[ -\int i\omega(k,\eta) d\eta\right].
\label{velocityexpansion}
\end{equation}

We can insert equations (\ref{expand}) and (\ref{velocityexpansion}) back in
equation (\ref{LBexpansion}). Solving the equations for
the frequency to second order in $\tilde
\lambda$ yields, 
\begin{equation}
\omega(k) = c_s k - i \gamma k^2 \tilde \lambda,
\end{equation}
where
\begin{equation}
c_s^2=\frac 13\frac 1{1+R};
\end{equation}
\begin{equation}
\gamma =\frac{R^2+4\left( 1+R\right) /5}{6(1+R)^2}.
\end{equation}
The inclusion of Rayleigh scattering has altered the dispersion relation
by reducing $\tilde \lambda$ due to the second term in equation (\ref{eq:lambda}).

The temperature perturbation is defined by
$\delta T/T\equiv\Theta(\nu,k,\mu)=-f^{(1)}(\nu,k,\mu)/\frac{\partial f^{(0)}(\nu)}{\partial\ln\nu}$ and can be expanded in Legendre polynomials
$\Theta(\nu ,k,\mu )=\sum_{l=0}\Theta_l(\nu ,k)(2l+1)P_l\left(\mu \right)$.

Thus, from equation (\ref{eq:Boltzmann}), we have the evolution of temperature
perturbation:
\begin{eqnarray}
\dot\Theta+ik\mu(\Theta+\Psi)=-\dot\Phi+\frac{a}{a_0}
\frac{1}{\lambda[(1+z(\eta))\nu]}
[\Theta_0-\Theta+\frac{1}{2}\Theta_2 P_2(\mu)+\mu u_b].
\label{eq:evoltemp}
\end{eqnarray}

If we ignore the angular dependence of Compton scattering,
equation (\ref{eq:evoltemp}) has the solution:
\begin{eqnarray}
(\Theta+\Psi)(\nu,k,\mu,\eta_0)=\int_0^{\eta_0} 
[-(\Theta_0+\Psi+\mu u_b)\dot\tau(\nu,\eta,\eta_0)
-\dot\Phi+\dot\Psi]e^{-\tau(\nu,\eta,\eta_0)}
e^{ik\mu(\eta-\eta_0)}d\eta,
\label{eq:solutemp}
\end{eqnarray}
where $\eta_0$ is the present epoch and $\tau(\nu,\eta_1,\eta_2)
\equiv\int_{\eta_1}^{\eta_2} \frac{a}{a_0}\frac{1}{\lambda[(1+z(\eta))\nu]}d\eta$.

Taking the multipole moments of equation (\ref{eq:solutemp})
and setting $u_b=3\Theta_1$, 
we have for $l\ge 2$ \citep{HS95},
\begin{eqnarray}
\Theta_l(\nu,k,\eta_0)\approx(\Theta_0+\Psi)(\eta_*)(-i)^l j_l(k\Delta\eta_*)
+\frac{3}{2l+1}(-i)^{l-1}\Theta_1(\eta_*)
[lj_{l-1}(k\Delta\eta_*)
\nonumber \\
-(l+1)j_{l+1}(k\Delta\eta_*)]
+(-i)^l\int_{\eta_*}^{\eta_0}(\dot\Psi-\dot\Phi)j_l(k\Delta\eta)d\eta,
\label{eq:multipole}
\end{eqnarray}
where $\Delta\eta\equiv\eta_0-\eta$, $\Delta\eta_*\equiv\eta_0-\eta_*(\nu)$, 
$\eta_0$ is the present epoch and 
$\eta_*(\nu)$ is the epoch at the last scattering. 
The fluctuations on the last scattering surface are obtained from 
$(\Theta_0+\Psi)(\eta_*)=(\hat\Theta_0+\Psi)(\eta_*)D(k,\nu)$ and
$\Theta_1(\eta_*)=\hat\Theta_1(\eta_*)D(k,\nu)$, where $D(k,\nu)$
is the damping factor (see its definition in \S 3.1). 
The undamped WKB solution 
$\hat\Theta_0(\eta)$ and $\hat\Theta_1(\eta)$, 
as well as the potential $\Psi(\eta)$ and $\Phi(\eta)$, can be obtained
from the Appendix in \citet{HS95}. They are all assumed not to be 
affected by Rayleigh scattering here, which is confirmed to be reasonable
by the consistency between our analytical and numerical results in \S 4.

Integrating over all $k$ modes of the perturbation, we have the power
spectrum of the auto-correlation function: 
\begin{equation}
\frac{1}{4\pi}C_l(\nu)=\frac{V}{2\pi^2}\int\frac{dk}{k}k^3|\Theta_l(\nu,k,\eta_0)|^2.
\end{equation}

\section{Observable Effects}

Rayleigh scattering produces several observable effects. \ We begin our
discussion with effects that are detectable in the Rayleigh-Jeans part of
the spectrum and then turn to effects that are detectable only in the Wien
tail.

\subsection{Suppression of Silk Damping}

Our treatment of Silk damping follows the analytical approach developed by
\citet{HS95}. At a given wavenumber, the amplitude of Silk damping
is the integrated photon diffusion length:
\begin{equation}
k_D^{-2}(\eta )=\int_0^\eta \gamma \tilde \lambda(\eta ^{\prime })d\eta
^{\prime }.
\end{equation}
Since Rayleigh scattering reduces the photon-diffusion length, it reduces the
amplitude of Silk damping. The damping factor at a given wavenumber is
weighted by the photon visibility function:
\begin{equation}
D(k,\nu )\equiv\int_0^{\eta _0}d\eta {\cal V}(\nu,\eta) \exp\left[-\frac{k^2}{k_D^2\left( \eta \right) }\right],
\end{equation}
where the ${\cal V}(\eta)$ is the photon visibility function
\begin{equation}
{\cal V}(\nu,\eta)\equiv -\dot\tau(\nu, \eta, \eta_0)\exp[ -\tau (\nu,\eta )],
\end{equation}
where $\tau $ is the photon mean free path (Figure 1). Figure 2 shows the 
damping
factor as a function of $k$ with and without Rayleigh scattering for a flat
vacuum-dominated universe with $\Omega _m=0.35,h=0.65$ and $\Omega_b=0.05$
(where $\Omega_m$ is the mass density in units of the critical density).

At low frequencies, the effect is small as photons decouple
before Rayleigh scattering is important.
At higher frequencies, the visibility function is
shifted to lower redshifts
where Silk damping is more important.  This  change
in the damping factor produces a proportional change in
the multipole spectrum at $l \sim 6000 k$.
The right panels in Figures \ref{fig:cll} and \ref{fig:clnu} show the amplitude
of the changes produced by this effect.

\begin{figure}
\epsscale{0.8}
\plotone{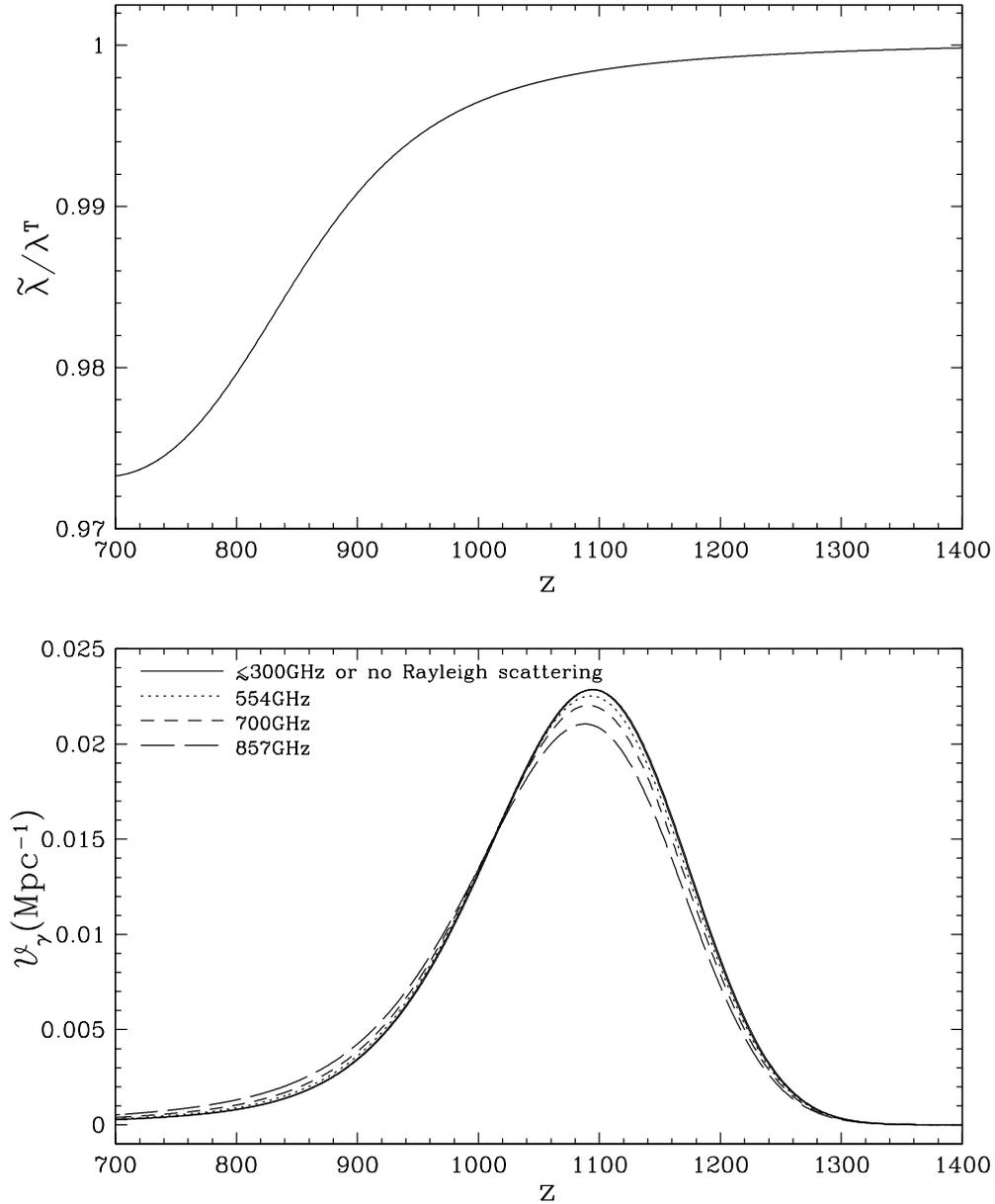}
\caption{The upper panel shows the ratio of the effective photon mean free path
with Rayleigh scattering ($\tilde\lambda$) to the photon mean free path without
Rayleigh scattering ($\lambda^{\rm T}$).
The lower panel shows that the photon visibility function $V(\nu,\eta)$
shifts toward the present epoch with increasing frequencies. }
\label{fig:visi}
\end{figure}

\begin{figure}
\epsscale{1.0}
\plotone{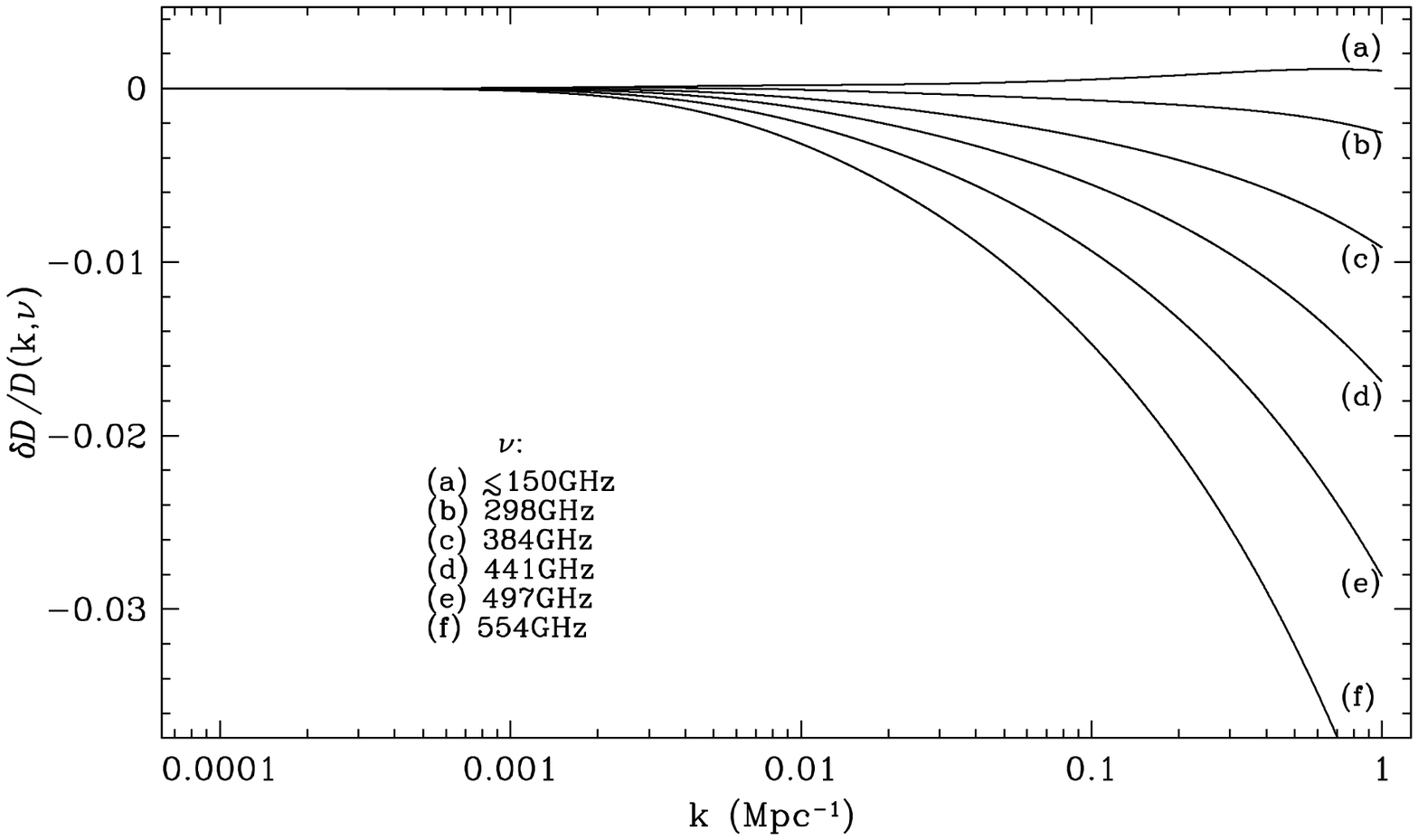}
\caption{Differences of damping factors: 
$\delta D/D(k,\nu)=D^{RS+T}(k,\nu)/D^T(k,\nu)-1$. The superscripts
$^{RS+T}$ and $^T$ stand for the cases with and without Rayleigh scattering,
respectively. At low frequencies, the increase of
$D^{RS+T}(k,\nu)$ comes mainly from the decrease of mean photon-diffusion 
length and the suppression of Silk damping; 
while at high frequencies, the decrease of
$D^{RS+T}(k,\nu)$ is mainly caused by the shift of visibility functions.}
\label{fig:damp}
\end{figure}

\subsection{Frequency Dependent Surface of Last Scattering}

Since the photon cross-section is frequency-dependent, the location of the
surface of last scattering will depend on frequency.  Since photons
emitted at lower redshift are emitted from regions with significant
amounts of Silk damping, they contribute little to the observed
temperature fluctuations.  Because of this latter effect,
the surface of last scattering is also $k$ dependent:
\begin{equation}
\left\langle 1+z\left( k,\nu \right) \right\rangle =D(k,\nu
)^{-1}\int_0^{\eta _0}d\eta\frac{a_0}{a}{\cal V}(\nu,\eta) \exp
\left[-\frac{k^2}{k_D^2\left( \eta \right) }\right].
\end{equation}

\subsection{Frequency Dependent Acoustic Wave Phase} 

Including Rayleigh scattering, the sound horizon at the last scattering 
$r^{RS+T}_s[\eta_*(k,\nu)]=\int_0^{\eta_*(k,\nu)} c_s d\eta$ will be frequency 
dependent. The higher the frequency, the farther the sound horizon.
It will shift the peak location by 
$\delta k/k(\nu)=1-r_s^{RS+T}(\eta_*)/r_s^T(\eta_*)$,
where the superscripts $^{RS+T}$ and $^T$ stand for the cases with and without 
Rayleigh scattering respectively.
Thus, the peaks in anisotropy spectra will 
shift in the direction of decreasing $l$: $\delta l/l=\delta k/k$ and
$\frac{\delta C_l}{C_l}=\frac{\partial \ln C_l}{\partial \ln l}
\frac{\delta l}{l}$. The higher the frequencies are, the more the peaks shift.

\subsection{Frequency Dependent Polarization Amplitude}

The amplitude of the microwave background polarization is proportional to
the photon mean free path at the surface of last scattering \citep{Z}. 
Since the photon mean free path depends upon frequency, the
amplitude of the polarization is now frequency dependent:

\begin{equation}
P\left( k,\nu \right) \propto \left\langle \lambda(\nu ,\eta
)\right\rangle =D(k,\nu)^{-1}\int_0^{\eta _0}d\eta {\cal V (\nu, \eta)} \exp \left[
-\frac{k^2}{k_D^2\left( \eta \right) }%
\right] \lambda(\nu ,\eta ).  \label{polarization}
\end{equation}
In equation (\ref{polarization}), we have weighted the photon mean free path
by the visibility function.

\section{Numerical Integration}

\begin{figure}
\epsscale{1.0}
\plotone{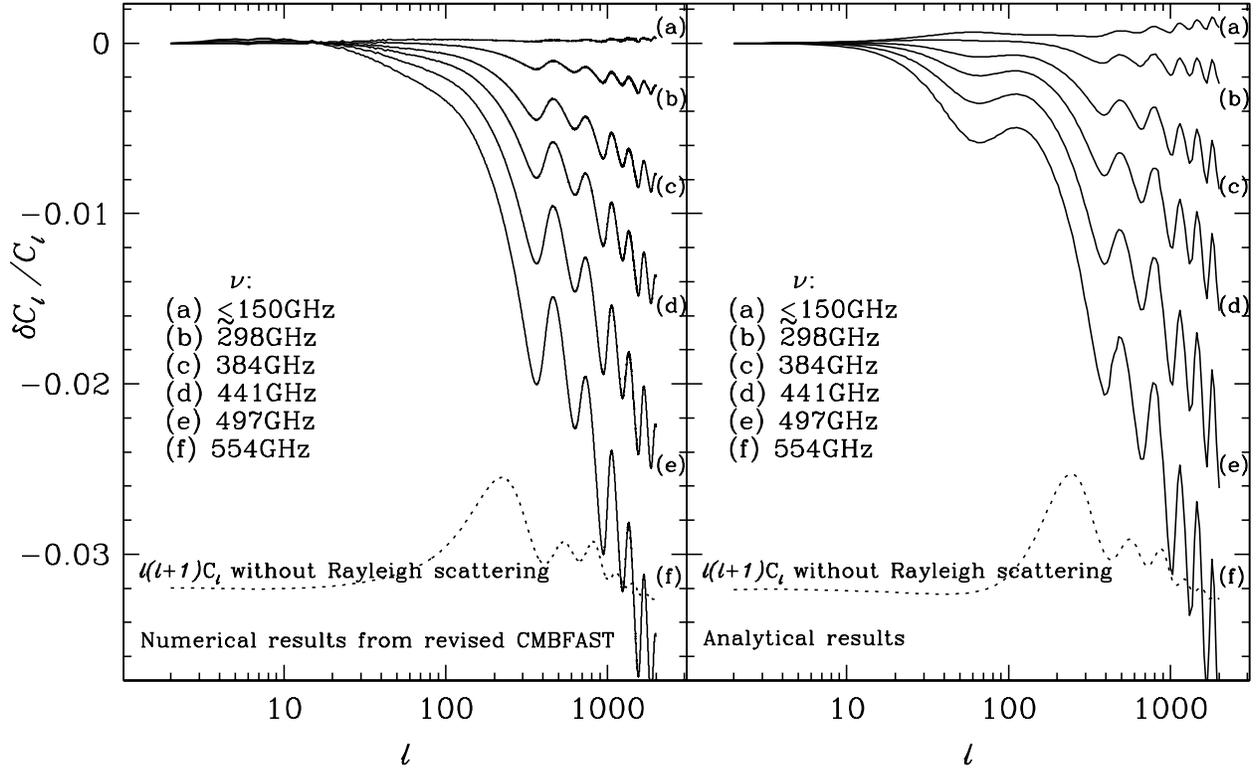}
\caption{Relative differences of anisotropy spectra with and without 
Rayleigh scattering: $\delta C_l/C_l(\nu)=C_l^{RS+T}(\nu)/C_l^T-1$
(solid curves). Basically, they keep the forms of their damping factor 
differences (see Fig. \ref{fig:damp}).  
The dotted curves represent the power spectra $l(l+1)C_l$ 
without Rayleigh scattering (in arbitrary unit). 
The oscillations of $\delta C_l/C_l$ show that the power spectra with Rayleigh 
scattering shift in the direction of decreasing $l$. 
The higher the frequencies are, the more the spectra shift 
(higher oscillation peak height). Note that the peak locations are not exactly
same in the numerical results and in the analytical results.
}
\label{fig:cll}
\end{figure}

\begin{figure}
\epsscale{1.0}
\plotone{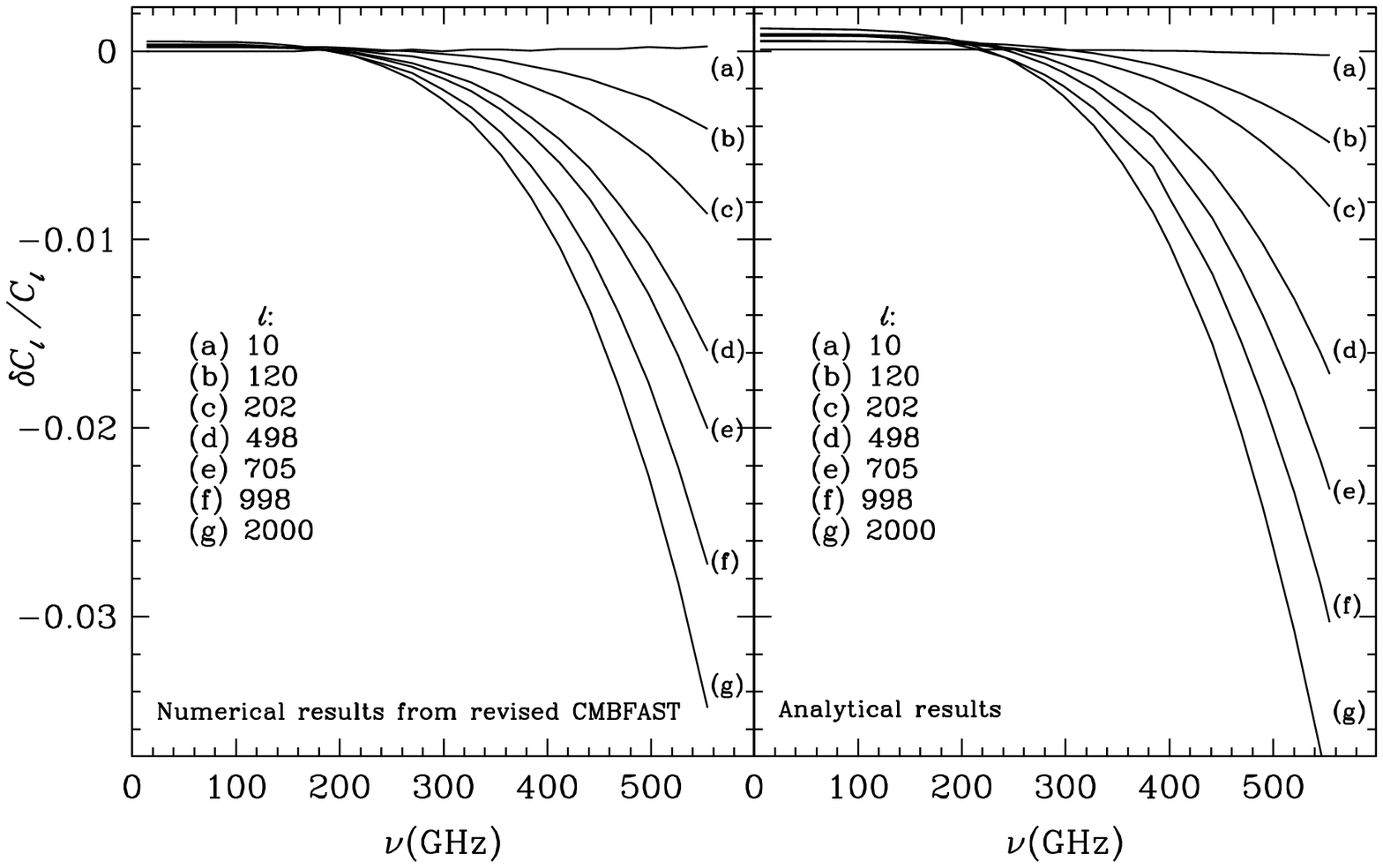}
\caption{Relative differences of anisotropy spectra with and without 
Rayleigh scattering: $\delta C_l/C_l(\nu)=C_l^{RS+T}(\nu)/C_l^T-1$. 
The positive differences at low frequencies come from the decrease of 
mean photon-diffusion length. The differences between the numerical results
and the analytical results are partly caused by slightly different peak
locations (see Fig.~\ref{fig:cll}).}
\label{fig:clnu}
\end{figure}

In this section, we describe results of numerical simulations of the
full Boltzmann equations.  
We have revised the CMBFAST\footnote{CMBFAST is available at:
http://physics.nyu.edu/matiasz/CMBFAST/cmbfast.html.} 
code \citep{ZS}
so that it can evolve the Boltzmann equations for frequency
dependent scattering.  The photon distribution function is
approximated by 50 different bins.  The fluctuations at each
frequency are evolved through the hierarchy of coupled equations
\citep{WS,BE,ZS}:
\begin{eqnarray}
\dot \Delta_{T0,j}^{(S)} &=& -k \Delta_{T1,j}^{(S)} + \dot \phi, \\
\dot \Delta_{T1,j}^{(S)} &=& {k\over 3} \left[\Delta_{T0,j}^{(S)}
-2 \Delta_{T2,j}^{(S)} + \psi\right] + \dot \kappa_j \left({v_b \over 3} - 
\Delta_{T1,j}^{(S)}\right), \\
\dot \Delta_{T2,j}^{(S)} &=& {k \over 5}\left[2 \Delta_{T1,j}^{(S)} -
3 \Delta_{T3,j}^{(S)}\right] + \dot \kappa_j\left[{\Pi \over 10}
- \Delta_{T2,j}^{(S)}\right], \\
\dot \Delta_{Tl,j}^{(S)} &=& {k \over 2 l+1} \left[l \Delta_{T(l-1),j}^{(S)}
- (l+1) \Delta_{T(l+1),j}^{(S)}\right]
- \dot \kappa_j \Delta_{Tl,j}^{(S)},\ \ {\rm for}\ \ l > 2,
\end{eqnarray}
where $\dot\kappa_j$(=$-\dot\tau$) is the (now frequency dependent) differential
optical depth and \\
$\Delta_{Tl,j}$(=$\Theta_l(\nu_j,k)/(-i)^l, l\ge0$) is the
$l$-th multipole moment of the temperature perturbation at frequency $\nu_j$.
All other symbols and equations (e.g. $\phi=-\Phi$, $\psi=\Psi$ and
$v_b=-iu_b$) are as defined in \citet{ZS}.

For the baryons, the momentum equation is now
modified to include the frequency-dependent coupling to the photons:
in \citet{ZS} is modified:
\begin{equation}
\dot v_b = {-\dot a \over a}v_b + c_b^2 k\delta_b + {4 \over 3 \rho_b}
\sum_j f_j \dot \kappa_j (3 \Delta^{(S)}_{T1,j} -v_b) + k\psi,
\end{equation}
where $f_j$ is the weight per bin. 
These modifications significantly slow CMBFAST, since it now
has to evolve 50 times more variables.  We have confirmed
that this modified code reproduces standard results.

In our numerical calculations,
we use the recombination results of $n_e$ and $n_H$ from 
RECFAST code to calculate the opacity to photon propagation \citep{SSS99}. 

Our numerical calculations agree remarkably well with the
analytical estimates of  (\S 2).
Figures \ref{fig:cll} and \ref{fig:clnu} show their relative differences as 
a function of $l$ and $\nu$.
As seen from the figures, the analytical results are qualitatively
consistent with the 
numerical results.
At high multipoles and high frequencies, the anisotropy spectra in both results
decrease because the visibility functions shift to lower redshifts
where Silk damping is more important.
The effect is up to 0.5\% for $\nu\sim350$GHz and 3\%
for $\nu\sim550$GHz at $l\sim1000$. 
At low frequencies and high multipoles, both the analytical and numerical 
results show that the anisotropy spectra with Rayleigh scattering are higher
($\sim$0.1\%) than the spectrum without Rayleigh scattering, 
which is caused by the increase of damping factor $D(k,\nu)$ at low frequencies 
(Fig.\ref{fig:damp}). At low $l$, that difference is too small to be 
physically significant because the cosmic variance uncertainty is comparably 
too large ($\sim$10\% for $l<10$).
In Fig.\ref{fig:cll}, comparing to the peaks of the anisotropy spectrum
without Rayleigh scattering, all the oscillation peaks of the spectrum
differences shift $\sim$ 1/4 ``period'' in the direction of decreasing
$l$, which supports our analyses about the phase shift caused by
frequency-dependent sound horizons at the last scattering in \S 3.3.
The oscillation amplitude of the spectrum difference is up to 0.5\% at
$\nu\sim550$GHz and $l\sim 1000$. 

There are small difference between
the numerical calculations and our
analytical estimates.  Note that the shift of the first peak
is more evident in the analytical results.
These differences may be due to 
the assumption that $\hat\Theta_0$, $\hat\Theta_1$
and the potential are not affected by Rayleigh scattering, or they may be due to
using equation (\ref{eq:multipole}) rather than
the exact solution of equation (\ref{eq:solutemp}).

\section{Discussion and Conclusion}

In this paper, we estimate the effects of Rayleigh scattering
on the microwave background. For {\it MAP}, Rayleigh scattering is a relatively
minor correction.  The largest effect is that it reduces the amount of 
Silk damping by $\sim 0.1\%$ at $\nu\sim 100$GHz and $l\sim 1000$.
The systematic error is comparable to {\it MAP}'s statistical errors.
For the {\it Planck Surveyor}'s highest frequencies, Rayleigh scattering
can have significant effects on the shape and amplitude of
the microwave background fluctuation spectrum.  
Overall, it reduces amplitudes, $C_l$, increasingly with increasing 
multipole number $l$, and, for observations at a single frequency, might
be confounded with varying $\Omega_bh^2$. However the frequency dependence
of Rayleigh scattering imparts a distinctive signature, so that the
effects pointed at in this paper should be easily corrected.
At $\sim$ 550 GHz,
the anisotropy spectrum decreases as much as 3\%. The peak shift between
the spectra of 100 and 500GHz makes their relative difference oscillate
as a function of $l$. The oscillation amplitude is as high as 0.5\%,
much larger than the statistical uncertainties in the {\it Planck Surveyor}
data.
While the high-frequency channels with $\nu>353$GHz are going to be used 
primarily for dust modeling, it
is useful to note that the cosmic microwave background fluctuations will not be
frequency independent.  By looking at regions of low dust emission,
the {\it Planck Surveyor} should be able to detect this effect.  This detection
would be an additional check of the recombination history of the universe,
and a direct measurement of the formation of atoms in the early universe.

\section{Acknowledgements}

We thank Wayne Hu, Arthur Kosowsky,
Sarah Seager and Matias Zaldarriaga for useful discussions.
We thank Uros Seljak and Matias Zaldarriaga for the
use of the CMBFAST code.
We thank Bruce Draine for pointing out an error in an early manuscript
and leading us to the correction.  DNS is supported in part
by the {\it MAP} program.

\end{document}